# Plasma activated PDMS microstructured pattern with collagen for improved myoblast cell guidance

Nikola Slepičková Kasálková [1], Veronika Juřicová [1], Dominik Fajstavr [1], Bára Frýdlová [1], Silvie Rimpelová [2], Václav Švorčík [1], Petr Slepička [1,*]

[1] Department of Solid State Engineering, The University of Chemistry and Technology Prague, 166 28 Prague, Czech Republic
[2] Department of Biochemistry and Microbiology, The University of Chemistry and Technology Prague, 166 28 Prague, Czech Republic
* Correspondence: petr.slepicka@vscht.cz

**Abstract:** We focused on polydimethylsiloxane (PDMS) as a substrate for replication, micropatterning, and construction of biologically active surfaces. The novelty of this study is based on the combination of argon plasma exposure of micropatterned PDMS scaffold, where the plasma served as a strong tool for subsequent grafting of collagen coating and their application as cell growth scaffolds, where the standard has been significantly exceeded. As part of scaffold design, templates with a patterned microstructure of different dimensions (50 x 50, 50 x 20 and 30 x 30 μm) were created by photolithography followed by pattern replication on a PDMS polymer substrate. Subsequently, the prepared microstructured PDMS replicas were coated with a type I collagen layer. The sample preparation was followed by the characterization of material surface properties through various analytical techniques, including scanning electron microscopy (SEM), energy-dispersive X-ray spectroscopy (EDS), and X-ray photoelectron spectroscopy (XPS). To evaluate the biocompatibility of the produced samples, we conducted studies on the interactions between selected polymer replicas with micro- and nanostructures and mammalian cells. Specifically, we utilized mouse myoblasts (C2C12) and our results demonstrate that we achieved excellent cell alignment in conjunction with the development of a cytocompatible surface. Consequently, the outcomes of this research contribute to an enhanced comprehension of surface properties and interactions between structured polymers and mammalian cells. The use of periodic micro- structures holds the potential for advancing the creation of novel materials and scaffolds in tissue engineering. These materials exhibit exceptional biocompatibility and possess the capacity to promote cell adhesion and growth.

## 1. Introduction

Newly developed materials are frequently designed to mimic the physicochemical properties of natural materials. Furthermore, the objective is for the biomaterial to function as the extracellular matrix (ECM), which governs tissue structure and cell phenotype at the microscale level [1]. In this context, conducting cytocompatibility testing both *in vivo* and *in vitro* using 2D substrates or 3D scaffolds is of paramount importance. Experiments at the *in vitro* level are extensively employed due to their ease and speed of execution, while also avoiding significant ethical concerns [2]. However, it is important to note that, for risk assessment, only a limited number of *in vitro* tests have been validated, primarily for individual chemical substances, and not for biomaterials or medical devices [3]. This standard also introduced new guidelines aimed at enhancing *in vivo* biocompatibility assessment, thereby promoting the reduction of the burden on animals and economic resources [4]. Tissue engineering (TE) represents an interdisciplinary field of biomedicine that owes its rapid advancement primarily to scientific progress in developmental biology and biomaterial sciences. [5-12]. In this context, the term scaffold refers to a carrier for the support and growth of cell cultures aiming to mimic their natural environment. The choice of cells from the spectrum found in the human body depends on their intended location and function such as fibroblasts and keratinocytes for skin reconstruction, osteoblasts for bone regeneration, or chondrocytes for cartilage replacements, among others [7-12]. These cells are already specialized and can be obtained directly from the organism (primary cells) or from a cell bank (secondary cells). The latter option includes undifferentiated stem cells, which possess a unique capacity for self-regeneration and differentiation. We classify stem cells as totipotent, pluripotent, and multipotent [13]. However, in clinical fields, the mentioned types are also referred to by other names. For example, in clinical contexts, stem cells are often called progenitor and precursor cells [14, 15].

Advances in precursor and stem cell biology, coupled with increasing interest in bioactive materials, have paved the way for the successful regeneration of potentially any tissue in the body [16]. Over the past 35 years of global research efforts, various artificially synthesized tissues have been successfully produced for non-clinical purposes [17-20]. For instance, modified small intestine organoids that have served in drug screening, the study of intestinal diseases, and intestinal physiology [21], can be named. Artificial human skin, composed of one to two cell types, has also found successful integration into clinical applications [22-24]. Nevertheless, achieving functional tissues with a wider range of cell types and complex anatomical structures has remained a significant challenge. This can be mainly caused by the obstacles associated with the traditional top-down TE technology approach, which relies on cell cultures and scaffolds for subsequent reconstruction [25].

Therapeutic approaches in TE must be tailored to the specific characteristics of the target tissue. To achieve this specificity, parameters such as surface topography and physicochemical properties, including surface wettability and surface charge, should be considered [26]. In the literature, a considerable number of studies on how surface chemistry and topography affect the adsorption of ECM proteins and the presentation of cell adhesion ligands can be found. Also, as early as 1994 Dr. Ikada already emphasized that the surface of a material has the most critical effect on its biocompatibility [27]. It is therefore not surprising that cell adhesion plays a fundamental role in in determining the subsequent behavior of cells such as their growth, proliferation, migration, and differentiation.

A wide range of techniques are available for synthesizing polymer nanostructures, which are generally categorized as "top-down" and "bottom-up" approaches. In the former, existing objects are reduced in size, thus, transforming microstructures into nanostructures. These techniques are based on different processes such as photolithography, ablation, and etching [28]. Laser modification allows the creating of nanostructures and can be applied on a wide range of substrates, very often polymers [29-39].

Conversely, the bottom-up approach generates novel structures by synthesis from smaller basic building units such as atoms and molecules. The so-called self-assembly is an example of a bottom-up technique that facilitates creating of nanostructures in a natural way using interactions between molecules and their self-organization into desired structures [40], including techniques like Chemical Vapour Deposition (CVD) [41]. However, it is worth noting that CVD can also be perceived as a top-down technique when considering that the resulting layer is formed on an already existing substrate.

Due to the extremely small dimensions of nanostructured materials, their characterization typically demands specialized equipment and techniques [42]. Photolithography entered the field of TE primarily due to the necessity for creating precise and reproducible structures on biomaterials serving as supports for cell growth and subsequent tissue reconstruction. A wide variety of natural and synthetic polymers can be used for this purpose often cross-linked utilizing light. Examples include the synthetic polymer poly(ethylene glycol) (PEG) and the natural polymer hyaluronic acid, which can be further modified with photoreactive groups [43]. Many studies published till now have been carried out focusing on the use of both photolithography, stereolithography, and soft lithography for the creation of scaffolds and biomaterials for the reconstruction of different types of tissues [44-47]. The relevant photomasks were created based on digital designs, taking into account already published studies [48-52].

Biological responses to immobilized microscale and nanoscale surface topographies, especially how the cellular responses are influenced by biochemical and biomechanical interactions with the extracellular matrix (ECM) was studied by Goering et al. [53]. Suitable scaffold structures and mechanical loading are essential for functional tendon engineering. The bipolar fibril structure of native tendon collagen is yet to be recaptured in engineered tendons [54]. High-throughput studies of collagen, chitosan, and collagen-chitosan hybrid biomembranes were carried out to characterize and compare key properties as a function of the applied hydrodynamic conditions during gelation. Specifically, depending on the biopolymer material used, varying flow conditions during biomembrane gelation caused width, uniformity, and swelling ratio to be differently affected and controllable [55]. PDMS with optimized stiffness provided enhanced bovine CEC response with higher density monolayers and increased phenotypic marker expression. This biomimetic approach demonstrates a successful platform to improve in vitro cell substrate properties of PDMS for corneal applications, suggesting an alternative environment for microfluidics and organ-on-chip applications [56]. Topographical patterns are a powerful tool to study directional migration. Grooved substrates have been extensively used as in vitro models of aligned extracellular matrix fibers because they induce cell elongation, alignment, and migration through a phenomenon known as contact guidance. This process, which involves the orientation of focal adhesions, F-actin, and microtubule cytoskeleton along the direction of the grooves has been primarily studied on hard materials of non-physiological stiffness [57]. Rapid and simple method to fabricate bundles of collagen type I, whose average thickness may be varied between about 4 μm and 9 μm dependent upon diluent temperature and ionic strength was presented in [58]. The durability and versatility of the collagen bundles was demonstrated with their incorporation into two in vitro models where the thickness and alignment of the collagen bundles resembled various in vivo arrangements. Rotational alignment of collagen gels was used to study the differences in contact guidance between mesenchymal and amoeboid cells. The collagen stiffness was increased through glycation, resulting in decreased MDA-MB-231 directionality in aligned collagen gels [59]. A platform which served as an in vitro testbed to assess neuro-regenerative potential of ASCs in aligned collagen fiber scaffolds and may provide guidance on next-generation nerve repair scaffold design was introduced in [60]. The most recent applications of soft materials in the modulation of cellular behavior, for tissue engineering and regenerative medicine, in drug delivery and for phototherapies was presented in [61]. Understanding how endothelial cell phenotype is affected by topography could improve the design of new tools for tissue engineering as many tissue engineering approaches make use of topography-mediated cell stimulation [62].

Poly(dimethylsiloxane) [PDMS] is an extremely versatile elastomer with excellent properties such as high optical transparency in a wide wavelength range (240–1,100 nm) [63], inertness, thermal stability, electrical and thermal insulation, high gas permeability, low dielectric constant, weak autofluorescence and very low surface tension (20.4 mN/m) [64]. Due to the lower temperature $T_g$ of this polymer, which is approximately of −123 °C, its properties are less sensitive to temperature than other rubbers [65]. The excellent elasticity of PDMS ($E \approx 1–3$ MPa) [66] and low price are also worth mentioning. The chemical structure of PDMS is unique since its backbone is made of silicon and oxygen instead of carbon. Silicon is a much bulkier atom than carbon, which enables PDMS to have a wider range of molar and molecular weights compared to polymers of the same dimension but with a carbon backbone. In addition, silicon is less electronegative than carbon, which also gives this polymer excellent hydrophobic properties. Hydrophobicity (contact angle with water of ~110°) [67] and bioinertness of this polymer results in very low adhesion, which may be undesirable for some applications, not only in TE, and methods of surface modifications or other surface functionalization must frequently be used [68-72]. In the field of medicine and biomaterials, PDMS is used in catheters [73], pacemakers [74], drug delivery [75], as part of some vascular graft systems [76], or as breast and cochlear implants [77, 78]. PDMS has also gained great popularity in the production of microchannels and microfluidic devices, which can be used, for example, to reconstruct blood vessels [79], vascularized tissues [80], or for rapid diagnostics [81]. The plasma treatment of PDMS with different types of plasma have been successfully used in [82], where effect of the type of plasma on the polydimethylsiloxane/collagen composites adhesive properties on PDMS film. Surface modifications to polydimethylsiloxane substrate for stabilizing prolonged bone marrow stromal cell culture were described in [83]. These modifications may find use also in photonic devices [84] or PDMS microdevice for biomedical applications [85].

In this study, we designed a microstructure that faithfully imitates the branched architecture of vascular and capillary networks. The combination of argon plasma activation of PDMS microstructures with collagen coating exhibits outstanding results for cell guidance of mouse myoblast cells. The summarizing Table I show the combination of PDMS surface treatment and its application in tissue engineering.

**Table I**. Selected treatments of PDMS materials and their use in tissue engineering.

| Type of modification | Application | Ref. |
| --- | --- | --- |
| Plasma treatment | Microfluidic applications | 67,68 |
| PE CVD | Applications that require stable hydrophilic surface property in PDMS | 69 |
| Photografting of PEG | Protein adsorption and cell adhesion | 71 |
| Imidazolium antimicrobial compound | Catheter fabrication | 73 |
| UV-light curing | Drug delivery | 75 |
| Soft lithography | Replication of cardiovascular flow conditions | 80 |
| Bulk oxygen, nitrogen or argon + collagen | Adhesive free skin substitutes | 82 |
| Bulk plasma treatment and collagen | Bone marrow derived stromal cells | 83 |
| Hot embossing | Photonic devices | 85-87 |
| Different bulk/ surface treatments | Biomedical applications | 88,89 |
| Surface plasma treatment and collagen | Mesenchymal stem cells | 90 |
| Different bulk/ surface treatments | Wettability and recyclability applications | 91-93 |
| Plasma / collagen treatment | Tissue engineering – different cell types | 94-96, 100, 101 |
| Bulk modification | Cell mechanobiology in muscle and nerve | 102 |

The promising results of this study suggest the potential for reconstructing three-dimensional vascularized tissue suitable for implantation. Photolithographic masks were designed to emulate the shape, size, and orientation of muscle fibers. To assess biocompatibility, the resulting microstructures were populated with muscle cells capable of dif-ferentiating into myotubes, thus forming functional muscle tissue. This innovative approach holds promise for the future production of scaffolds intended for the reconstruction of muscle tissue in patients with damaged or lost muscle tissue. We believe that this paper can bring an interesting new information for the field of polymers, targeted surface modification of polymeric materials. The outcomes of this research contribute to an enhanced comprehension of surface properties and interactions between structured polymers and mammalian cells. The use of periodic micro- and nanostructures holds the potential for advancing the creation of novel materials and scaffolds in tissue engineering. These materials exhibit exceptional biocompatibility and possess the capacity to promote cell adhesion, growth, and differentiation and will serve as a strong basis for possible application of these surfaces in tissue engineering mainly in the field of cytocompatibility enhancement and cell guidance.

## 2. Results and Discussion

### 2.1 Surface morphology of the replicas

Scanning electron microscopy (SEM) was used to monitor and assess the success of photoresist pattern replication on the PDMS substrate. Atomic force microscopy (AFM) analysis was not suitable since the dimensions of these periodic patterns are on the microscale. To study the surface morphology, PDMS samples prepared by replication from photoresist patterns with the dimensions of 50 x 50 and 30 x 30 μm were selected.

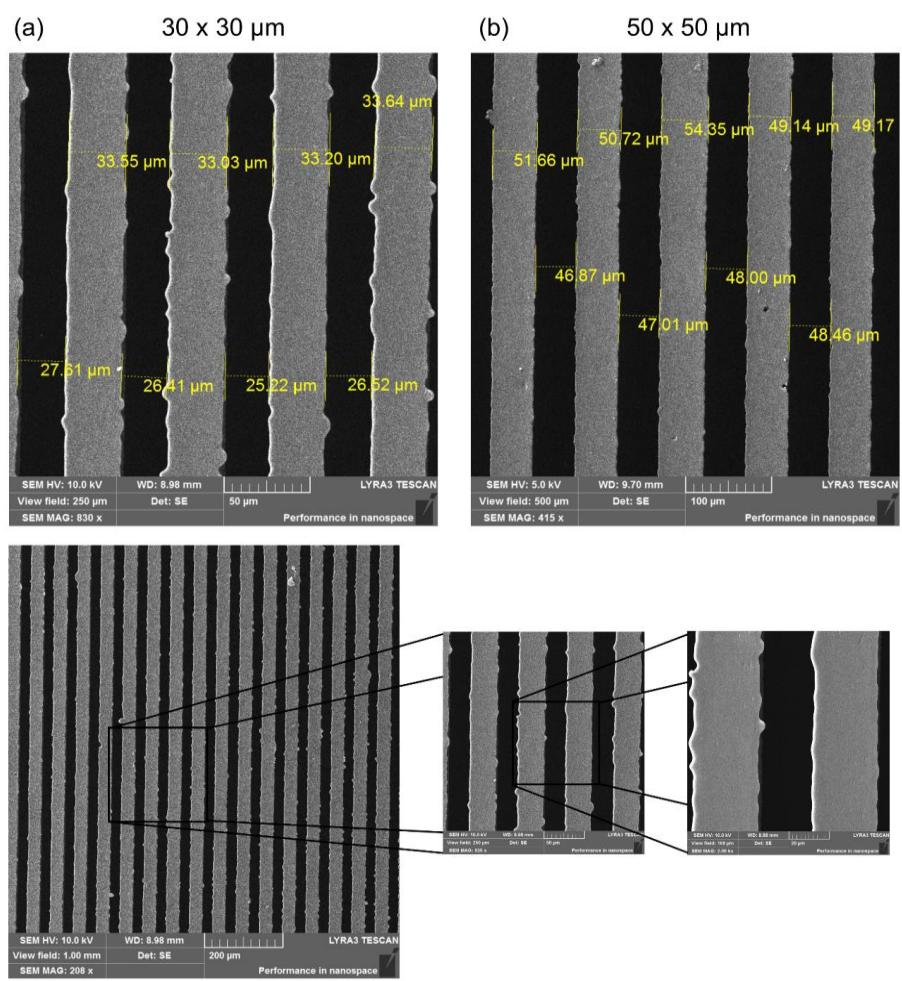

**Figure 1:** SEM images of PDMS negative replicas. (a) 30 x 30 μm and (b) 50 x 50 μm photoresist pattern. Detailed SEM image of PDMS negative replica (30 x 30 μm) demonstrating the pattern edges.

The images from the SEM analysis (Figure 1) confirmed the preservation of the original dimensions of the photoresist pattern gaps/grooves, which were monitored using confocal microscopy (as introduced in the Experimental section). More detailed SEM images reveal small defects on the 30 x 30 μm PDMS replica (Figure 1, bottom image), which may already occurred during the photoresist pattern production process or casting. These defects can be caused by several reasons, such as insufficient or too long exposure time, or bad removal of the photoresist when a residual layer remains [86, 87]. Tiny defects were observed to a much lesser extent on the 50 x 50 μm sample, indicating that their formation had already occurred before PDMS casting.

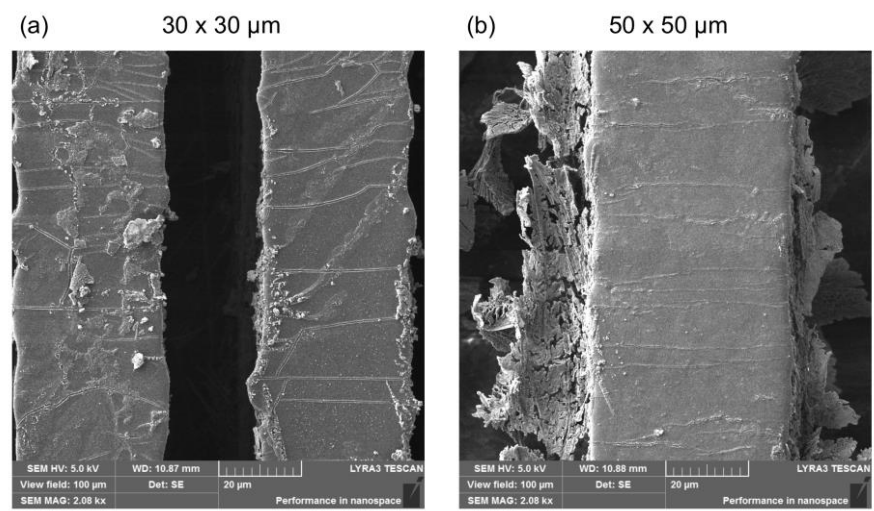

**Figure 2:** Detail SEM images of PDMS negative replicas coated with a type I collagen layer after lyophilization. (a) Pattern size 30 x 30 μm and (b) 50 x 50 μm.

To observe the morphological changes that might occur due to the coating of the surfaces, PDMS replicas coated with a type I collagen layer were subsequently scanned. In the detailed images, with the size of 100 × 100 μm², a change in their structure can be observed mainly in the grooves themselves with a disturbed discontinuous surface (Figure 2). Crystals of salts (e.g., NaCl, KCl) contained in the phosphate buffer (PBS), in which the samples were stored before lyophilization, are most likely also found on this surface. The presence of salts was also confirmed using energy dispersive spectroscopy (EDS), the results of which will be discussed in the following paragraph. In addition, delamination of the collagen coating can be also observed near the grooves in Figure 2b. Delamination could occur due to long sample storage (>14 days) before lyophilization or due to the lyophilization process itself, which can lead to changes in the protein structure and its aggregation [88]. During freeze-drying, the removal of water can lead to collagen denaturation, which can result in breakage of hydrogen bonds and loss of some secondary and tertiary structures of the protein.

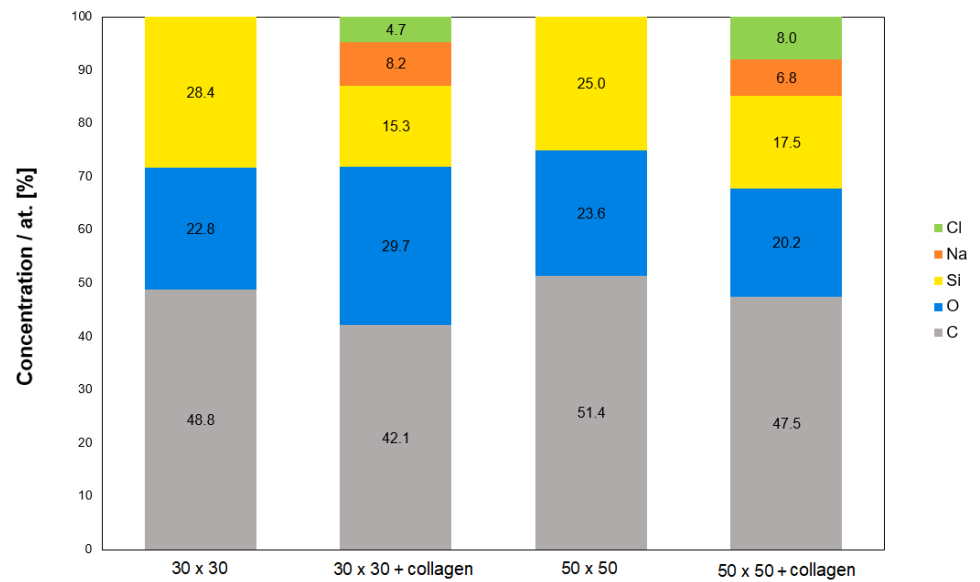

**Figure 3:** Chemical composition of PDMS surfaces (30 x 30 μm, 50 x 50 μm) before lyophilization and after collagen coating; EDS analysis, measured area size 30 × 30 μm².

*2.2 Study of surface chemistry*

Chemical changes occurring in collagen-coated PDMS after freeze-drying were monitored using energy dispersive spectrometry (EDS) (Figure 3) and using X-ray photoelectron spectroscopy (XPS) (Figure 4). After the lyophilization of PDMS itself, the EDS spectrum is not expected to change significantly because the elemental composition of PDMS is not changed by the lyophilization process. However, the thickness of the collagen coating corresponds to approximately 30 μm [89], which means that assuming a continuous layer, only collagen should be detected in the spectra of both analytical methods after lyophilization of the coated PDMS samples. However, the results of the analyses used do not directly indicate the presence of collagen (Figure 3).

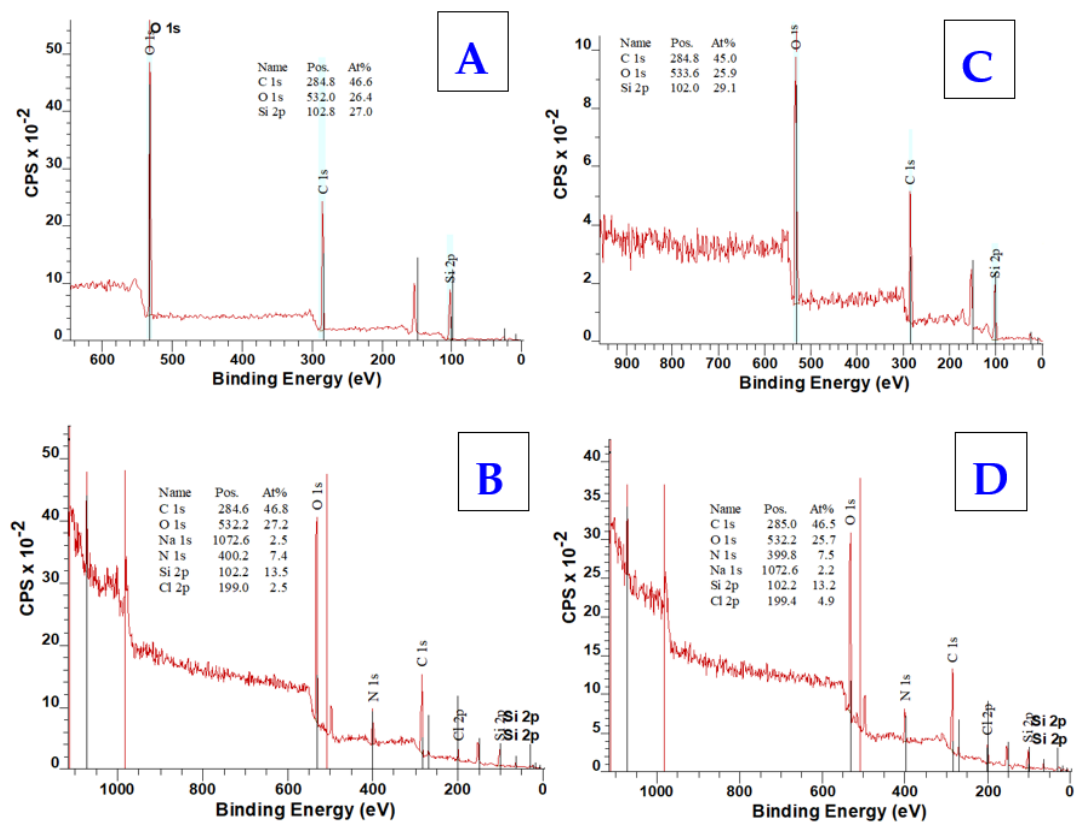

**Figure 4** XPS spectra of PDMS surfaces (30 x 30 μm) before lyophilization (A) and after collagen coating (B) acquired form XPS analysis, and XPS spectra of PDMS surfaces (50 x 50 μm) before lyophilization (C) and after collagen coating (D) acquired form XPS analysis.

Therefore, the confirmation we were looking for could be found mainly in the increase in carbon (C), oxygen (O), and nitrogen (N), which are the elements found in the collagen structure both from EDS and XPS analysis (Figure 4). The C and O elements are also contained in the PDMS chain and, therefore, their increase or decrease may not be a clear indicator. On the contrary, there should be a decrease in silicon (Si), which is only found in the PDMS structure. The decrease in the element Si was confirmed in both samples. Nitrogen detection by EDS analysis was too low (<2.4 wt.%, i.e. wt.%) and there were no significant changes (<1.5 wt.%). However, the results of both EDS and XPS analyses also confirmed the presence of salts, specifically the Na and Cl elements, the precipitation of which was also observed by SEM (Figure 2). The XPS spectrum further revealed information about the bond states and electron configurations of the given elements. In this spectrum, signals for Na and Cl with the corresponding electron configurations, i.e., Na in the stable configuration ($1s^2$, $2s^2$, $2p^6$) and Cl ($1s^2$, $2s^2$, $2p^6$, $3s^2$, $3p^6$) would be expected in the NaCl molecule. Although both of these assumptions were confirmed in the XPS spectrum, the signal for the chlorine atom (Cl) in the 2p region could be present for both sodium chloride (NaCl) and potassium chloride (KCl). Deconvoluted C1s spectra for particular samples are introduced in Figure 5. As it is obvious from the deconvoluted spectra, the results are in good agreement with those observed by Keffalinou et al. [90], some particular differences have to be attributed to the type of applied plasma, where we have used argon inert treatment. It is obvious, that after collagen coating there is no significant difference between both deconvoluted spectra (Figure 5 B and D), but the difference is visible compared to uncoated samples (Figure 5 A and C). This should support the evidence of additive collagen layer, which have been successfully added onto plasma treated PDMS microstructure. We have also determined the contact angle of selected samples. Pristine microstructure foil has contact angle 109°, plasma treated PDMS foil exhibited with contact angle 24°, and plasma treated PDMS microstructure with collagen had a contact angle 54°, the results also in good agreement with Keffalinou et al. [90]. Wettability-Patterned Surfaces Prepared by an Atmospheric-Pressure Plasma Jet were used in microfluidic devices [91], wettability have had also a significant impact for microchannel for enhancing droplet coalescence and demulsification [92] and surface activation of polymer blends for recyclability [93].

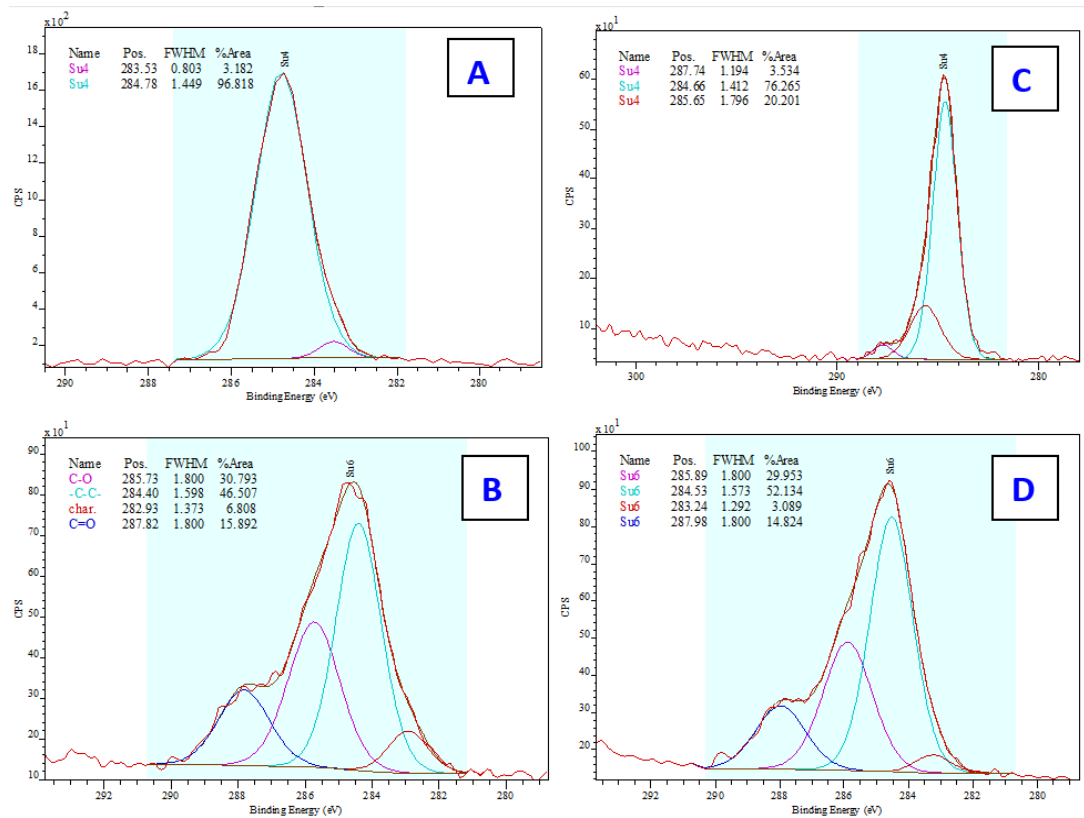

**Figure 5** Deconvoluted C1s XPS spectra of PDMS surfaces (30 x 30 μm) before lyophilization (A) and after collagen coating (B) acquired form XPS analysis, and XPS spectra of PDMS surfaces (50 x 50 μm) before lyophilization (C) and after collagen coating (D) acquired form XPS analysis.

## 2.3 Study of PDMS cytocompatibility

To study cytocompatibility of the prepared samples, PDMS replicas of photoresist patterns with the dimensions of 50 x 50 μm and 30 x 30 μm were selected, while TCPS was chosen as a control substrate. On these samples, adhesion (day 1) and proliferation (day 3) of mouse C2C12 myoblasts were studied. We have also attempt to determine the number of cells after 6 day from seeding, but the cells was so well spreaded, also in multiple layers, that we were not able to determine the cell number precisely, so we only introduced the images which show perfect cell alignment due to the prepared microgrooves.

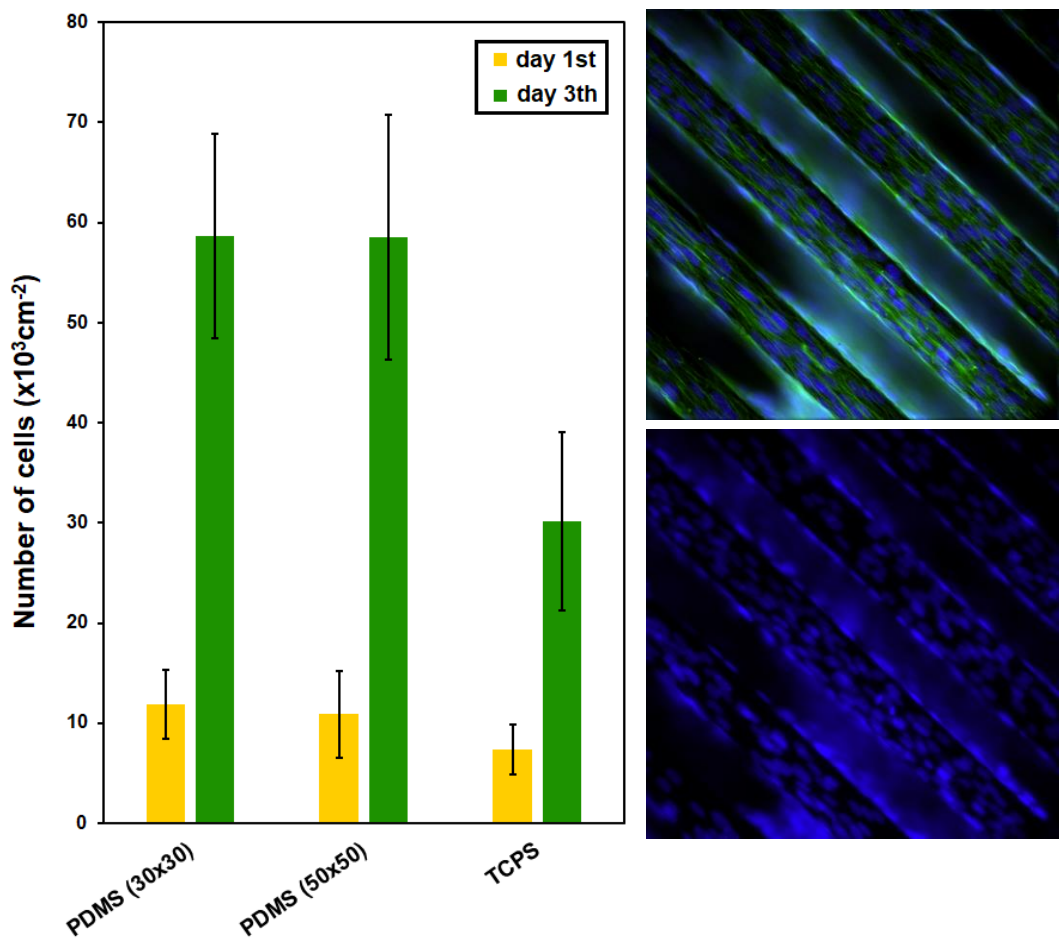

**Figure 6** Number of adhered (day 1) and proliferated (day 3) C2C12 cells cultured on PDMS surfaces (30 x 30 μm, 50 x 50 μm) after collagen coating, tissue polystyrene (TCPS) for comparison is also introduced. PDMS surface (line 50 x 50 μm) after collagen coating - fluorescence microscopy images of C2C12 cells with labelled cytoskeleton (in green) and nuclei (in blue) are also introduced.

From the results shown in Figure 6, it is apparent that the collagen-coated PDMS replicas significantly promoted cell adhesion and growth compared to the control, already after the first 24 h. This difference was even more pronounced in the following days of cultivation, when on day 3, the average number of cells growing on PDMS replica samples per cm$^2$ reached ca. 58,000. Major factor that influence the cell behavior is the immediate cell microenvironment. The effect of extracellular matrix (ECM) proteins controlling cell survival, proliferation, migration and directed differentiation was described for mesenchymal stem cells cells on collagen support [94] or neuronal cell lines [95]. Further, regarding C2C12 cell morphology, the most important result is their linear alignment due to the reorganization of the cell cytoskeleton along the grooves of the photoresist patterns, which was already manifested after the first day of their cultivation (see Figure 6 and 7). Both types of PDMS substrates studied significantly supported cell growth, while the difference between the dimensions of the patterns was not decisive for the success of cell orientation. Cells on PDMS micropattern with collagen on the 6$^{th}$ day from seeding are well spreaded, and aligned along the pattern. Similar studies have been done with microfluidic chambers by Keffalinou et al [90]. By plasma pre-treated and collagen-coated PDMS-based microfluidic chambers the have been able to enhance and stabilize mesenchymal stem cell growth inside. Simple surface engineering of polydimethylsiloxane with polydopamine for stabilized mesenchymal stem cell adhesion [96], PDMS bonding technologies for microfluidic applications were described precisely in [97].

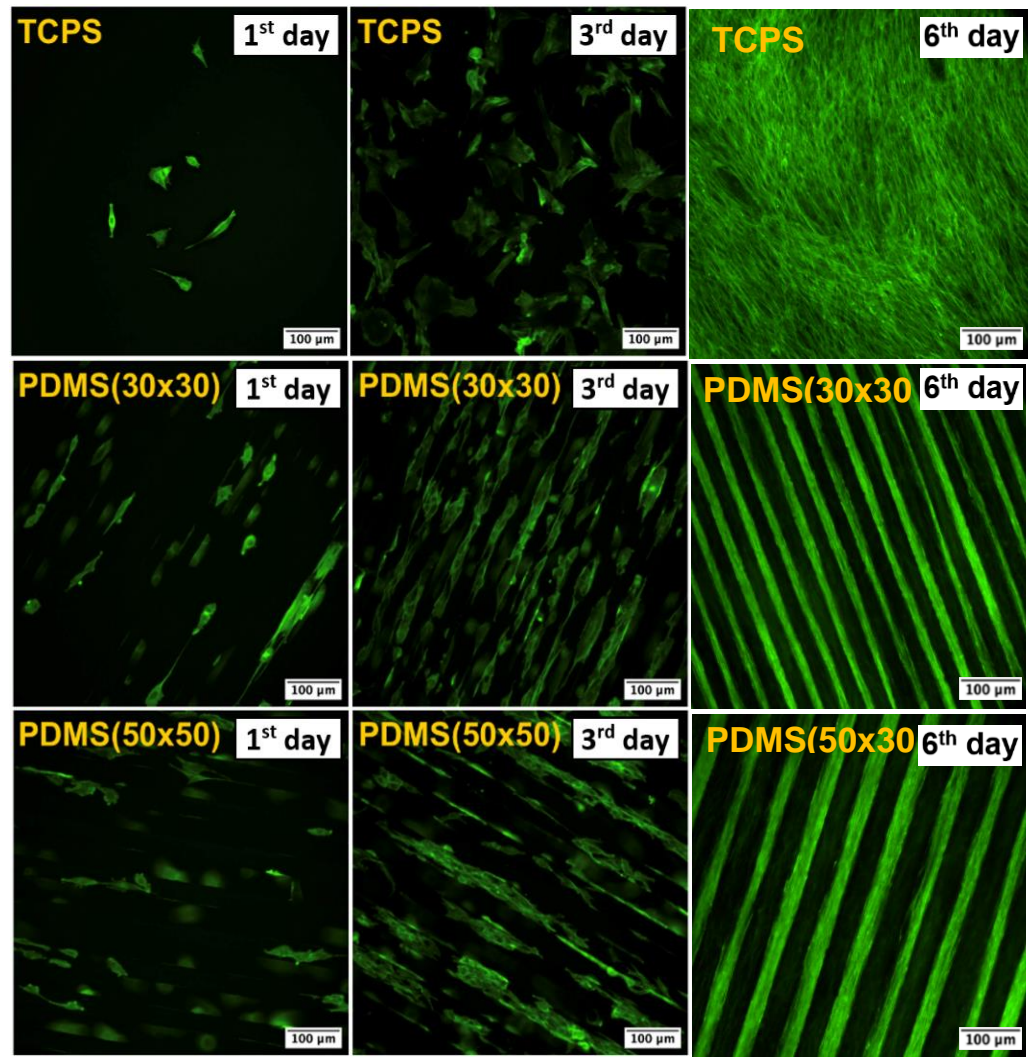

**Figure 7** Images from fluorescence microscopy images of C2C12 with phalloidin (Atto 488) stained cells, on PDMS surfaces (30 x 30 μm, 50 x 50 μm) after collagen coating, tissue polystyrene (TCPS), the results 1st day, 3rd day and 6th day from seeding are presented.

The behavior of myoblast cells was monitored on microstructured PDMS substrates (30 x 30 and 50 x 50 μm), which were covered with a layer of bovine collagen type I before the optimal cell density was applied. From the results in Figure 7, it is clear that cells preferentially populated grooves over ridges, which may be the reason for hindered cell proliferation due to reduced space for adhesion and growth. Preferential growth of different cell types in wells was also observed in several earlier studies that investigated microstructured substrates [98, 99]. However, in several places on the substrate, cells were also able to bridge depressions and attach to ridges. Similar behavior was shown, for example, by C2C12 mouse myoblasts on microstructured PDMS scaffolds with smaller pattern sizes (3–12 μm) [100]. Another important observed phenomenon was the direction of cell growth and stretching, which was identical to the direction of the pattern.

## 3. Materials and Methods

*3.1 Materials used in photolithography and replication*

*PDMS polymer*

Polydimethylsiloxane (PDMS) elastomer kit Sylgard 184® with the base/curing agent ratio 10:1 was purchased from Sigma Aldrich (Merck, USA).

*Phosphor bronze wafer*

The phosphor bronze wafer was milled using a Gravos GV21 CNC robotic micromill. A circle of 2.5 inch in the diameter was cut out to serve as a substrate for depositing the photoresist layer. For better adhesion and better optical properties of the photoresist, the bronze wheel was etched with a mixture of $HNO_3:H_2O$ in a ratio of 1:2. The wheel was then placed in an ultrasonic bath for 5 min. to remove grease and dirt. The surface was then rinsed with isopropanol and deionized water and enabled to dry in an oven.

*Photoresist (SU-8 3035)*

Photoresist SU-8 3035 from the German company Micro Resist Technologies enabled the creation of microstructures on a phosphor bronze wafer. The development of the photoresist is carried out by the mr-600 developer from the same manufacturer. The recall time varies between 4-5 min. Its great advantage is sufficient adhesion to the metal substrate without the need to apply an adhesive layer.

*3.2 Preparation of microstructured substrates*

A total of five types of microstructures were designed with different gap sizes ranging from 10 x 10 μm to 100 x 100 μm (width dimension x gap dimension), while the linear motif of the grooves always remained the same (see Figure S1). Photomasks were ordered for production by JD Photo Data; UK. The development of photoresist SU-8 3035 on a phosphor bronze wafer was experimentally carried out. The designed patterns constructed by photolithography were further analyzed with a laser confocal microscope LEXT OLS 5000. Three different dimensions – 50 x 50 μm, 50 x 20 μm, and 30 x 30 μm – were selected and evaluated from the digital microstructure designs. From the visual representation of the iQ-Analyzer-X software, it was possible to determine the real dimensions of the photoresist patterns (Figure S2 a-c), while their average values are summarized in Figure S2 at the bottom.

*3.3 Preparation of photoresist pattern replicas*

In the first step, taking into account the desired shape and size of the photoresist patterns, a PETG mold was designed, which was used for casting PDMS replicas. After the 3D printing was completed, the molds were properly washed and dried. The next step was the preparation of the PDMS mixture itself, which consists of two main components – a silicone elastomer and a catalyst. Correct component ratios are important to achieve optimal PDMS properties, while the recommended weight ratio of 10:1 (elastomer: catalyst) was followed. The prepared PDMS mixture was carefully cast into a PETG mold. To ensure better spreading of the PDMS mixture and to remove air bubbles, the mold was placed in a desiccator for 20–30 min. After desiccation, the molds were transferred to a laboratory oven at 80 °C for 60 min. to enable faster curing of the PDMS. After the PDMS mixture had solidified, the replicas of the photoresist patterns were removed from the PETG mold with gentle movements or with the help of a scalpel.

Sterilization of the PDMS replicas took place in a Tuttnauer ML series laboratory steam sterilizer for 20 min. The sterilization temperature was of 121 °C under an increased pressure of 210 kPa. This additional pressure is created by injecting steam into the autoclave chamber, which increases the temperature and pressure inside the chamber and enables the material to be sterilized. After reaching the desired sterilization time and temperature, the steam is discharged from the chamber and the pressure is again reduced to the atmospheric value.

*3.4 Plasma modification of replicas*

The surface of the PDMS replicas intended for cytocompatibility tests was modified with plasma on a table-top Sputter Coater, model SCD 050 from BAL-TEC in etching mode. Four samples were evenly distributed on the cathode in the chamber so that they did not overlap and their distance from the anode was 5 cm. Plasma exposure took place at a working pressure of 8 Pa. This process took place for 80 s at a power of 8 W. After the modification was completed, the samples had to be sterilized again. The chosen sterilization method this time was exposure to UV radiation for 20 min. on both sides to avoid undesired exposure of the samples to water vapor.

During the plasma modification of the PDMS surface with argon plasma, argon atoms, and ions collide with PDMS molecules on the surface of the material. These collisions can lead to the formation of active sites, such as radicals (e.g., methyl or peroxide radicals) and ions, which can react with surrounding molecules. These radicals can react with the surrounding PDMS molecules and lead to the formation of new functional groups, such as carbonyl (−C=O) and hydroxyl (−OH) groups (Figure S3). As a result of the formation of polar groups on the PDMS surface, its wettability also improves.

*3.5 Coating of functionalized replicas*

Plasma modification of the PDMS surface improves the ability of its surface to adhere to other biomolecules [101], such as, among others, collagen. The resulting active sites on the PDMS surface can further react with the amino groups in the collagen structure, thus, creating strong bonds between the two materials. Without prior functionalization of the PDMS surface with argon plasma, low adhesion of the collagen coating may occur, which increases the risk of its delamination or peeling. The coating of the PDMS surface with bovine type I collagen was performed in a sterile laminar flow cabinet. In addition to type I collagen, type IV collagen also exists. However, type IV collagen reduces the contact angle of PDMS with water by only 10°, which was found to be insufficient regarding water contact angle changes in previous studies [102]. A stock solution of bovine type I collagen (6 mg·mL$^{-1}$) was dissolved in acetic acid solution in phosphate-buffered saline (PBS; 2 mg·mL$^{-1}$ of $CH_3COOH$ solution) so that the final concentration of collagen for coating was 90 μg·mL$^{-1}$. The solution thus prepared was applied onto the surface of the replicas, after which they were incubated at room temperature (RT, 21 °C) for 1 h (Figure S4). Finally, the collagen solution was aspirated with a pipette and the samples were rinsed by PBS several times to wash away the unreacted collagen residues. PDMS replicas were stored in PBS at 4 °C before cell seeding. Samples intended for analysis by scanning electron microscopy were dehydrated to remove residual moisture by lyophilization for 24 h.

*3.6 Analytical methods*

The morphology of the sample surfaces was characterized using the scanning electron microscope FIB-SEM LYRA3 GMU (Tescan, Brno, Czech Republic). The acceleration voltage was set to 10 kV. To ensure the conductivity of the samples, their metallization was performed using the sputtering technique (Quorum Q300T) by deposition of a Pt layer (the thickness of 20 nm, Pt target, purity of 99.9995 %).

The elemental composition was measured by energy-dispersive X-ray spectroscopy (EDS, analyzer X-ManN, 20 mm² SDD detector, Oxford Instruments, United Kingdom), while the accelerating voltage for SEM-EDS analysis was set to 10 kV.

The elemental composition on the material surface was analyzed by X-ray photoelectron spectroscopy (XPS) using a spectrometer ESCAProbeP (Omicron Nanotechnology Ldt., Taunusstein, Germany). As a source, a monochromatic X-ray at an energy of 1486.7 eV was used. Atomic concentrations of elements were determined from the individual peak areas using CasaXPS software. The techniques of elemental analysis were used for the detection of changes induced by collagen grafting, however for more specific analysis some addition techniques may be used [103].

Wettability of the studied samples was determined by measurement of contact angles (CA, θ) on goniometer Advex Instruments (Brno, Czech Republic) connected to the SEE System 7.1 program. Analysis of CA was performed at room temperature with 8 μL drops of distilled water (dyed with methyl violet) using a Transferpette® automatic pipette (Brand, Wertheim, Germany) at 6 different positions of 3 samples in parallel and perpendicular direction. Subsequently, the drops were photographed and evaluated by 3 marked points.

*3.7 Cytocompatibility testing*

Before bioassays, the samples were sterilized for 30 min. in 70% (*v/v*) ethanol in water. Then, the samples were placed into 12-wells cell culture plates and carefully rinsed with PBS (1.5 mM, pH of 7.4) solution, which was then aspirated and changed for high-glucose Dulbecco's modified Eagle's medium with stable L-glutamine (DMEM, Thermo Fisher Scientific, USA) with 10% (*v/v*) fetal bovine serum (FBS, Thermo Fisher Scientific, USA). The thus prepared PDMS samples were then inoculated with mouse myoblasts (C2C12; 15,000 cells per cm²; three replicates, received from Sigma-Aldrich s.r.o.) so that the resulting volume in the well was 1 mL. The cell cultivation took place for 6 days at 37 °C, 95% humidity, and 5% $CO_2$. Tissue culture polystyrene (TCPS) was used as a control material. After 1, 3, and 6 days, C2C12 cells were rinsed with PBS (37 °C) and then fixed with 4% (*v/v*) formaldehyde in PBS for 20 min. in the dark. Subsequently, the fixative was removed, and the samples were washed twice with PBS and stained for 20 min. with a solution of phalloidin-Atto 488 (1 μg·mL$^{-1}$; Atto-TEC GmbH, Germany) and DAPI (0.3 μg·mL$^{-1}$; Merck, USA) in PBS in the dark. Then, the staining solution was removed, the samples were washed with PBS twice, and subjected to fluorescence microscopy (Olympus IX-81; EMCCD camera, Hamamatsu) using 10× objective (total magnification 100×) so that at least 10 fields of view from the entire sample were taken. FITC and DAPI filter cubes (Olympus, Japan) were used, and the images from individual channels were merged and background-corrected.

**4. Conclusions**

In this research study, we have designed and prepared PDMS scaffolds using templates with a patterned microstructure of varying dimensions. These templates were generated by using photolithography, and the resulting patterns were replicated onto a PDMS polymer substrate. Subsequently, the fabricated microstructured PDMS replicas were coated with a type I collagen layer and subjected to cytocompatibility testing using C2C12 mouse myoblasts. Notably, the PDMS photoresist replicas were found to significantly promote the growth of C2C12 cells, resulting in cell densities exceeding those observed on the control substrate (TCPS). Another advantage of these patterns lies in their ease of use for further biopolymer replication, such as for PLLA or PLGA biopolymers.

**Author Contributions: Author Contributions:**
Conceptualization, N.S.K. and V.S.;
Methodology, N.S.K., S.R., and D.F.;
Validation, N.S.K. and V.S.;
Formal analysis, S.R.;
Investigation, D.F., V.J. and S.R;
Data curation, D.F. and B.F;
Writing—original draft preparation, P.S., V.J.;
Writing—review and editing, P.S., N.S.K., and S.R.
Supervision, P.S.;
Funding acquisition, P.S.
All authors have read and agreed to the published version of the manuscript.

**Funding:** This work was supported by the grant GACR under project 22-04006S and project OP JAK ExRegMed, project number No CZ.02.01.01/00/22_008/0004562, of the Ministry of Education, Youth and Sports, which is co-funded by the European Union.

**Institutional Review Board Statement:** Not applicable.

**Informed Consent Statement:** Not applicable.

**Data Availability Statement:** All data are contained within this article.

**Conflicts of Interest:** The authors declare no conflict of interest.